\documentclass[12pt]{iopart}

\usepackage{iopams}
\usepackage[dvips]{graphicx}

\begin{document}

\title{On the possibility of $q$-scaling in high energy production processes}

\author{Maciej Rybczy\'nski, Zbigniew W\l odarczyk }
\address{Institute of Physics, Jan Kochanowski University,
\'Swi\c{e}tokrzyska 15, 25-406 Kielce, Poland}
\ead{Maciej.Rybczynski@ujk.edu.pl,
zbigniew.wlodarczyk@ujk.kielce.pl}
\author{Grzegorz Wilk}
\address{National Centre for Nuclear Research,
Ho\.{z}a 69, 00-681 Warsaw, Poland} \ead{wilk@fuw.edu.pl}

\begin{abstract}
It has recently been noticed that transverse momenta ($p_T$)
distributions observed in high energy production processes exhibit
remarkably universal scaling behaviour. This is the case when a
suitable variable replaces the usual $p_T$. On the other hand, it
is also widely known that transverse momentum distributions in
general follow a power-like Tsallis distribution, rather than an
exponential Boltzmann-Gibbs one, with a (generally energy
dependent) nonextensivity parameter $q$. Here we show that it is
possible to choose a suitable variable such that {\it all} the
data can be fitted by the {\it same} Tsallis distribution (with
the same, energy independent value of the $q$-parameter). They
thus exhibit $q$-scaling.
\end{abstract}


\maketitle

Almost fifty years ago Hagedorn developed a statistical
description of momentum spectra observed experimentally \cite{H1}.
It predicts an exponential decay of differential cross sections
\begin{equation}
E\frac{d^3\sigma}{d^3p} \simeq C\cdot \exp\left( -
\frac{p_T}{T}\right)\label{eq:H1}
\end{equation}
for transverse momenta, whereas in experiments one observes
non-exponential behaviour for large transverse momenta. Hagedorn
then proposed the 'QCD inspired' empirical formula describing the
data of the invariant cross section of hadrons as a function of
$p_T$ over a wide range \cite{H2}:
\begin{eqnarray}
  E\frac{d^3\sigma}{d^3p} = C\cdot \left( 1 + \frac{p_T}{p_0}\right)^{-\alpha}
  \longrightarrow
  \left\{
 \begin{array}{l}
  \exp\left(-\frac{\alpha p_T}{p_0}\right)\quad \, \, \, {\rm for}\ p_T \to 0, \smallskip\\
  \left(\frac{p_0}{p_T}\right)^{\alpha}\qquad \qquad{\rm for}\ p_T \to \infty,
 \end{array}
 \right .
 \label{eq:H2} .
\end{eqnarray}
with  $C$, $p_0$  and $\alpha$ being fit parameters. This becomes
pure exponential for small $p_T$ and pure power law for large
$p_T$\footnote{Actually this QCD inspired formula was proposed
earlier in \cite{CM,UA1l}.}.

When looking at  $p_T$  spectra (\ref{eq:H1}) of secondaries
produced in high energy multiparticle processes, it is commonly
assumed that the temperature $T$ of the hadronizing system, when
treated as a statistical ensemble, can be connected with the
observed mean transverse momentum $\langle p_T\rangle$. Usually,
however, the system is far from thermal equilibrium and the
individual event temperature $T$ cannot correspond to the mean
transverse momenta. The temperature fluctuates from event to event
(or also in the same event). Such a situation is described by a
nonextensive generalization of statistical mechanics proposed
quite some time ago \cite{Tsallis}. There is one new parameter,
$q$, in addition to the temperature $T$, and the main formula of
interest here is the Tsallis distribution,
\begin{equation}
h_q\left( p_T\right) = C_q\cdot\left[ 1 -
(1-q)\frac{p_T}{T}\right]^{\frac{1}{1-q}} \quad \stackrel{q
\rightarrow 1}{\Longrightarrow}\quad h\left( p_T\right) =
C_1\cdot\exp \left(-\frac{p_T}{T}\right),\label{eq:Tsallis}
\end{equation}
(where $C_q$ is a normalization constant). This coincides with Eq.
(\ref{eq:H2}) for
\begin{equation} \alpha
= \frac{1}{q - 1}\quad {\rm and}\quad p_0 = \frac{T}{q -
1}.\label{eq:coincides}
\end{equation}
This approach has been shown to be very successful in describing
very different physical systems \cite{Tsallis}. Among them are
also multiparticle production processes of a different kind (see
\cite{WW1,WW2} for recent reviews). The basic conceptual
difference between (\ref{eq:H2}) and (\ref{eq:Tsallis}) is in the
underlying physical picture. In (\ref{eq:H2}) the small $p_T$
region is governed by {\it soft physics} possibly described by
some unknown nonperturbative theory or model, and the large $p_T$
region is governed by {\it hard physics} believed to be described
by perturbative QCD. In (\ref{eq:Tsallis}), the nonextensive
formula is valid in the whole range of $p_T$ and it does not claim
to originate from any particular theory. It is just a
generalization of the usual statistical mechanics and merely
offers the kind of general unifying principle, namely the
existence of some kind of complicated equilibrium (or steady
state) involving all scales of $p_T$, which is described by two
parameters, $T$ and $q$. The temperature $T$ describes its mean
properties and the parameter $q$, known as the {\it nonextensivity
parameter}, describes action of the possible nontrivial long range
effects believed to be caused by fluctuations (but also by some
correlations or long memory effects) \cite{Tsallis}. In fact, it
was shown in \cite{qWW} that $q$ is directly connected to the
variance of $T$,
\begin{equation}
q = 1 + \frac{Var(T)}{<T>^2}, \label{eq:q}
\end{equation}
and therefore describes its intrinsic fluctuations. For $q = 1$
one recovers the usual BG distribution (Eq. (\ref{eq:H1})). In
other words, the widely used thermal bath concept fails to satisfy
conditions allowing us to introduce the notion of thermal
equilibrium in the BG sense: it is always finite and can hardly be
considered homogenous. In fact, in many cases it only occupies a
fraction of the allowed phase space or even has a fractal-like
structure. In such cases, a heat bath cannot be described by a
single parameter $T$. One has to extend the parameter space to
account for all effects mentiond above and  use Eq.
(\ref{eq:Tsallis}) with a new additional parameter
$q$\footnote{Cf. \cite{WW1,WW2} for further references concerning
specific applications of this approach to hadronic and nuclear
physics in last decade. Recent examples of power-like
distributions fitted by nonextensive Tsallis formula are provided
in \cite{qcompilation,CW} , by PHENIX \cite{PHENIX} and STAR
\cite{STAR} experiments at RHIC and CMS \cite{CMS}, ATLAS
\cite{ATLAS} and ALICE \cite{ALICE} experiments at LHC. Finally,
the possible QCD origin of such fluctuations and/or correlations
could probably be traced back to the nonperturbative QCD, for
example, \cite{QCD}.}.

\begin{figure}[h]
\includegraphics[width=0.5\textwidth]{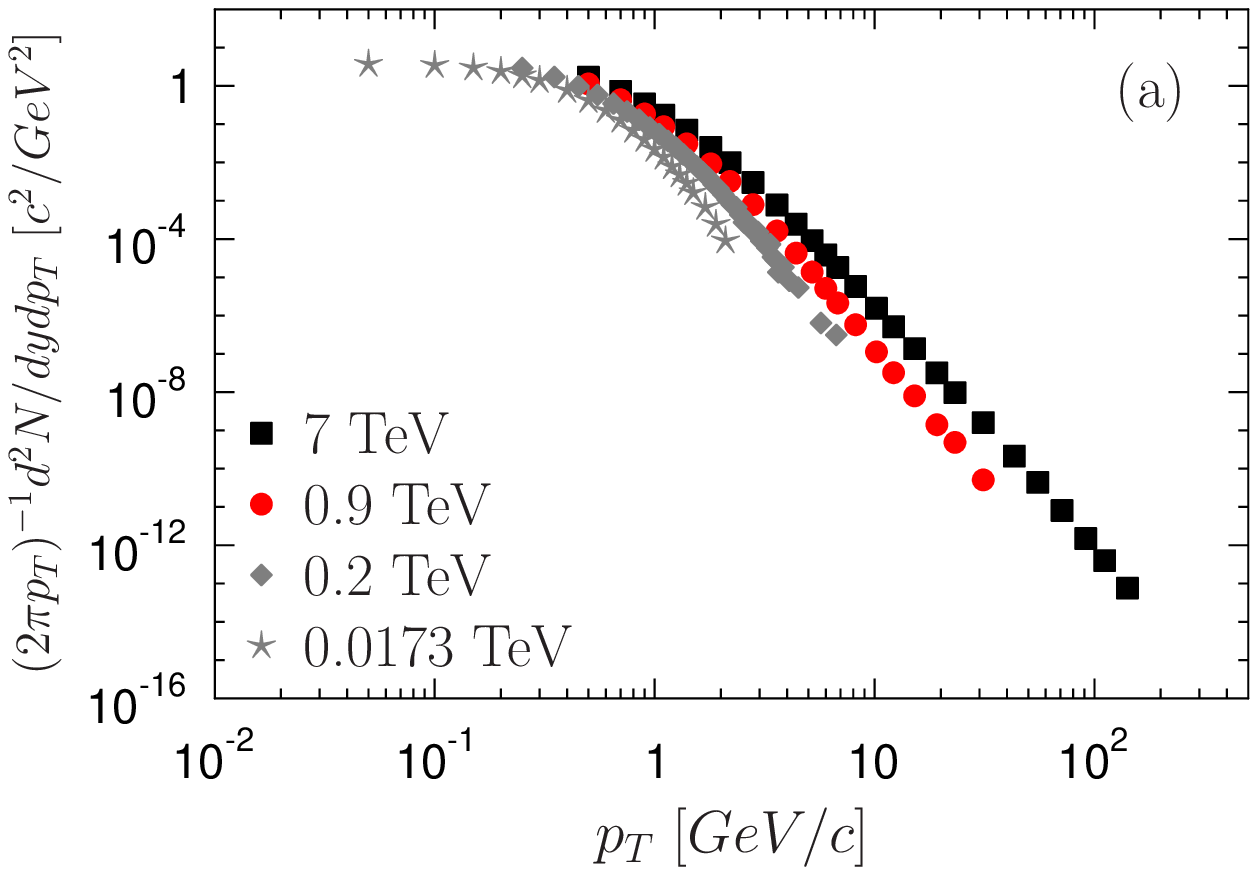}
\includegraphics[width=0.5\textwidth]{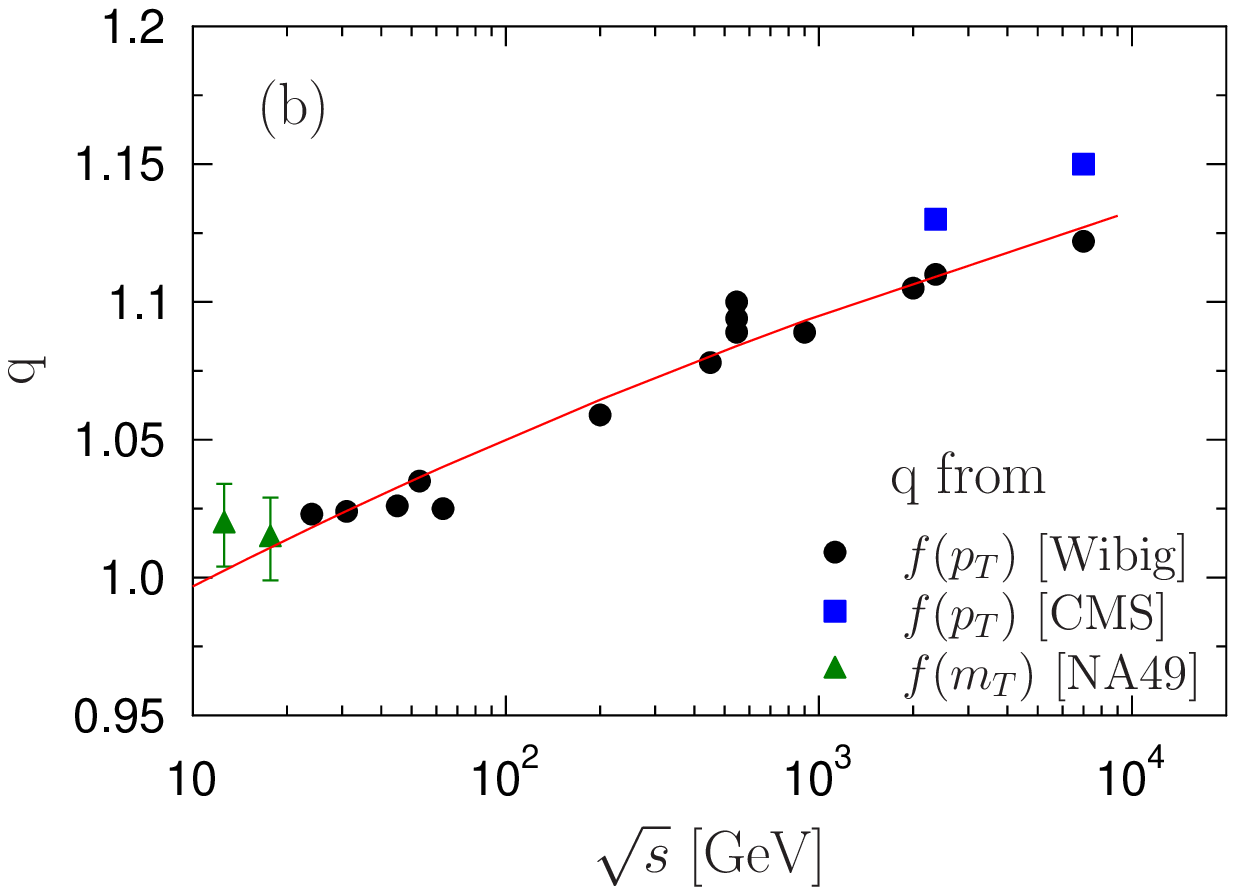}
\caption{(Color online)$(a)$ Transverse momenta distributions
considered by us. $(b)$ The corresponding values of the parameter
$q$ obtained from Tsallis fits. Data are from compilation by Wibig
\cite{Wibig}, NA49 \cite{NA49}, UA1 \cite{UA1} and CMS
\cite{CMS}.}\label{Fig_exp}
\end{figure}
\begin{figure}[h]
\includegraphics[width=0.5\textwidth]{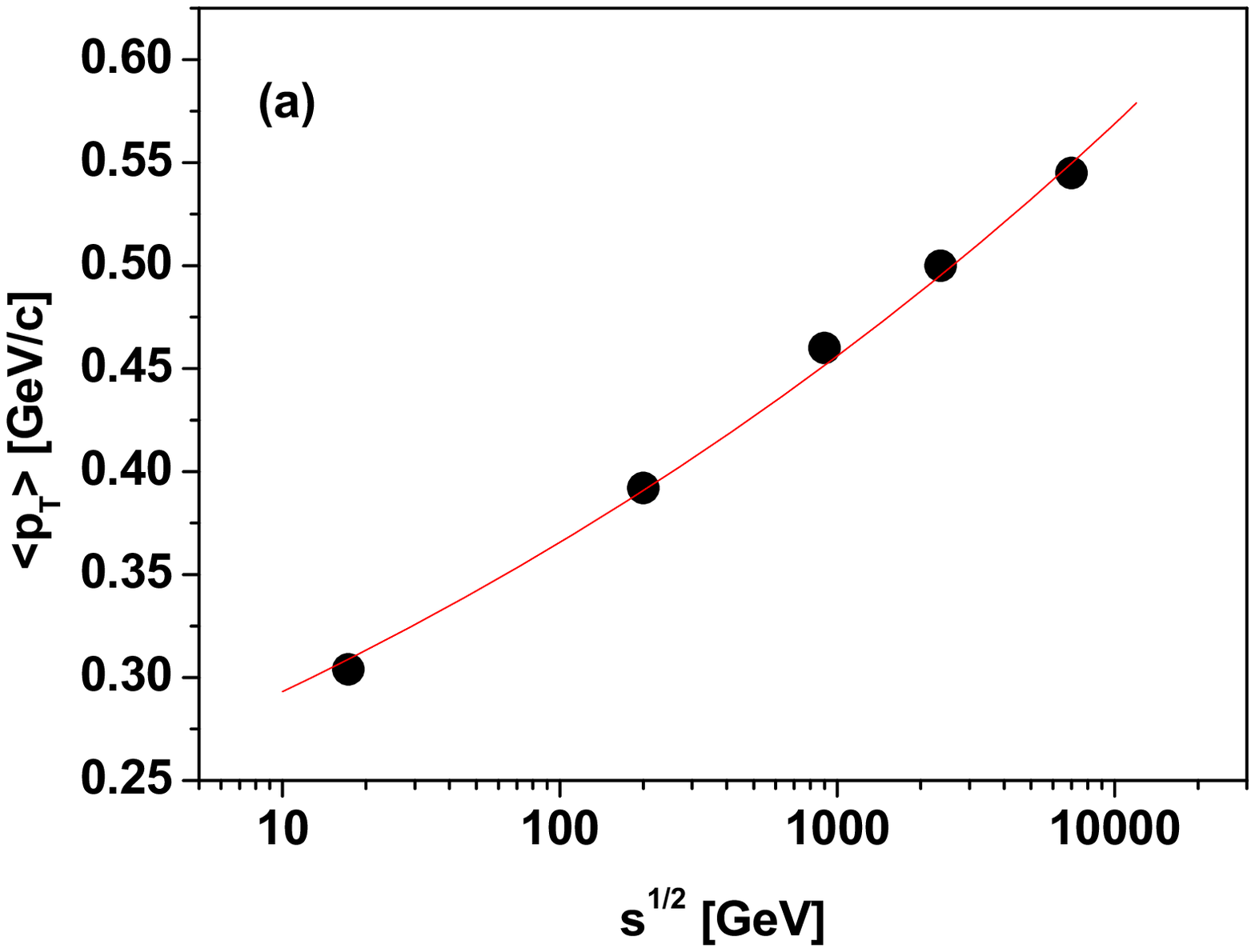}
\includegraphics[width=0.5\textwidth]{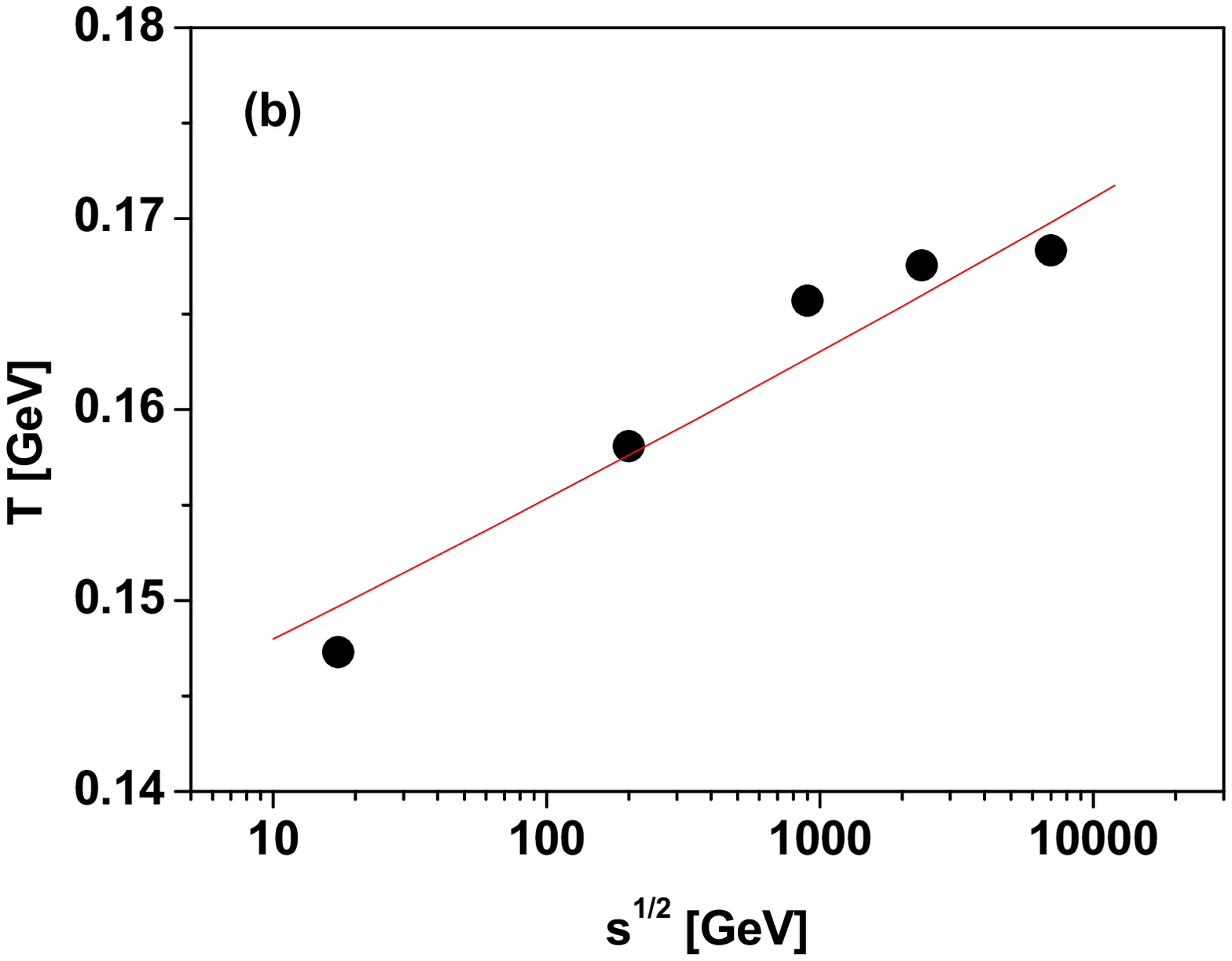}
\caption{(Color online) $(a)$ Experimental values of mean
transverse momenta, $\langle p_T(s)\rangle$, of charged particles
produced in $p+p$ and $p+\bar{p}$ collisions \cite{CMS,NA49,UA1}
(cf. Fig. \ref{Fig_exp}a for corresponding distributions at
selected energies). $(b)$ The resulting $T(s)$ as given by Eq.
(\ref{eq:p_TT}). }\label{Fig_ptT}
\end{figure}

In what follows we concentrate on data on $p_T$ distributions of
secondaries produced in $p+\bar{p}$ and $p+p$ collisions at
selected energies (covering the wide range of available energies
at roughly the same distance in logarithmic scale): $7$ TeV and
$0.9$ TeV from CMS \cite{CMS}, $200$ GeV from UA1 \cite{UA1} and
$17.3$ GeV from NA49 \cite{NA49}, cf. Fig. \ref{Fig_exp}a. They
can be fitted at each energy by using Eq. (\ref{eq:Tsallis}) with
$h(p_T)=dN/dp^2_T$. From these fits one finds values of $q$ for
different energies, $q(s)$ shown in Fig. \ref{Fig_exp}b. It can be
represented by $q(s) = 4/3 - 0.4\left(\sqrt{s}\right)^{-0.075}$
(full line). The $q(s)$ from Fig. \ref{Fig_exp}b can be translated
to energy dependence of the temperature, $T(s)$, with the help of
$\langle p_T(s)\rangle$,
\begin{equation}
\langle p_T\rangle = \frac{T}{4 - 3q}, \label{eq:p_TT}
\end{equation}
Using $\langle p_T(s)\rangle $ as evaluated experimentally, cf.
Fig. \ref{Fig_ptT}a (it can be parameterized by $\langle
p_T(s)\rangle = 0.235 \left( \sqrt{s} \right)^{0.096}$) one gets
the $T(s)$ presented in Fig. \ref{Fig_ptT}b, which can be
represented by $ T(s) = 0.141 \left(\sqrt{s}\right)^{0.021}$. To
summarize: all $p_T$ data considered here can be fitted with a
Tsallis formula, Eq. (\ref{eq:Tsallis}), by using an energy
dependent parameter $q(s)$ and $T(s)$.

Recently there have been attempts to use an energy independent $q$
to describe $dN/dp_T$ and to check for possible scaling behaviour
in $p_T$ \cite{BUB}. It was found that to this end one has to use
a $p_T$-dependent form of the nonextensivity parameter,
\begin{equation} q\left( p_T\right) = \frac{q_0 - \left(
q_0 - 1 \right) \theta\left( p_T\right)}{1 - \left( q_0 - 1\right)
\theta \left( p_T\right)}, \label{eq:Theta}
\end{equation}
where $q_0 = 1.12 \pm 0.06$ and $\theta \left( p_T\right) = 2 \log
\left[ \log \left( 1 + \kappa p_T\right) \right]$ with $\kappa =
0.013$ GeV$^{-1}$. It was shown that any experimentally accessible
$q$ depends on $p_T$ growing from $q = 1$ for $p_T = 0$ and
attaining $q = q_0$ for $p_T = 9/\kappa$.

The distributions $dN/dp_T$ shown in Fig. \ref{Fig_exp}a differ
for different energies. However, as shown in \cite{LMLMP,MP1,MP2},
one can find a single scaling function $F(\tau)$, independent of
energy, and a suitable scaling variable $\tau$ such that one
observes a scaling: $h(p_T,\sqrt{s})\rightarrow F\left( \tau =
f\left(p_T,\sqrt{s}\right) \right)$, analogous to Feynman or KNO
scaling \cite{KNO}. Prompted by {\it geometrical scaling}
behaviour found in deep inelastic scattering data, the following
universal variable has been proposed \cite{MP2},
\begin{equation}
\tau = \frac{p_T^2}{Q^2_{sat}},\qquad{\rm with}\qquad
Q^2_{sat}\left( p_T \right) = Q_0^2\left(
\frac{p_T}{W}\right)^{-\lambda}. \label{eq:tau}
\end{equation}
\begin{figure}[h]
\includegraphics[width=0.5\textwidth]{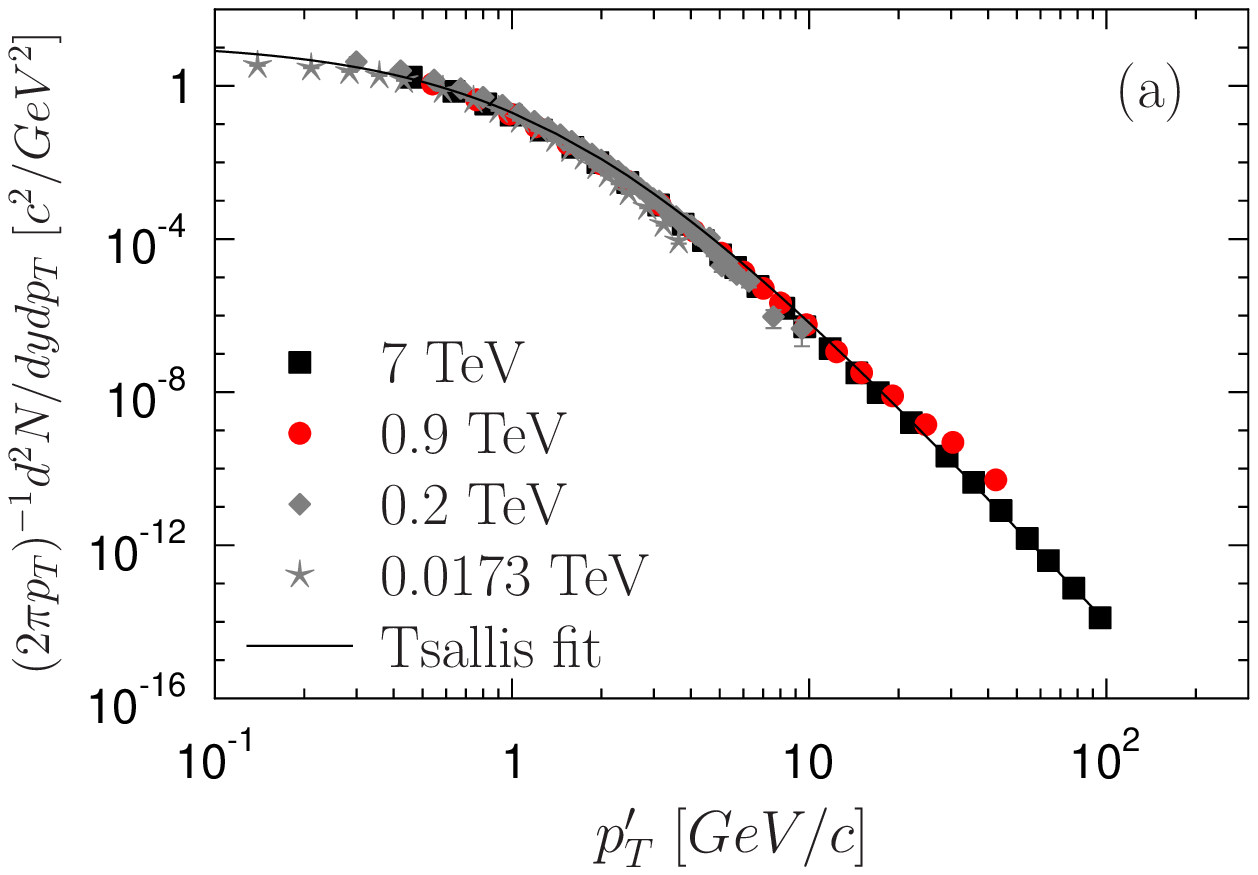}
\includegraphics[width=0.5\textwidth]{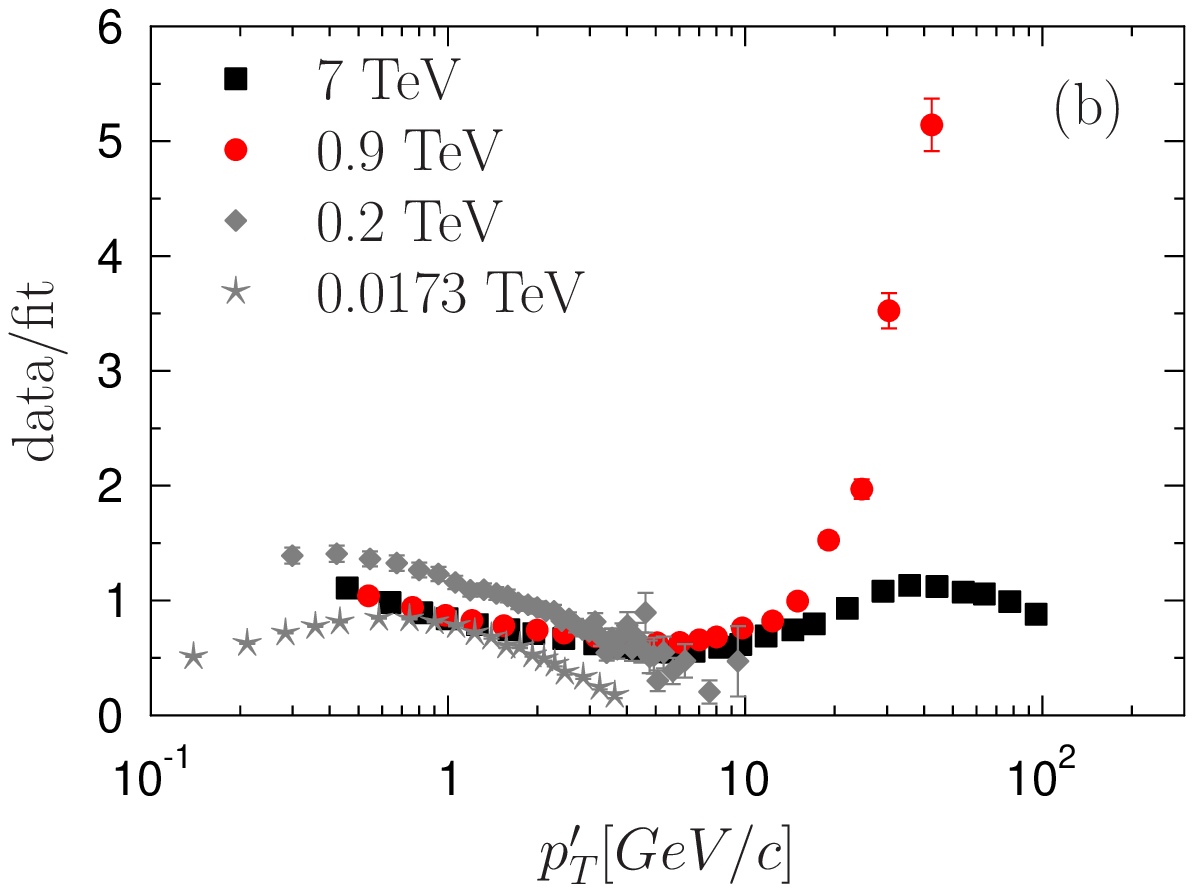}
\caption{(Color online) $(a)$ Data for transverse momentum
distributions for different energies \cite{NA49,UA1,CMS} plotted
by using the scaling variable $p'_T$ defined in \cite{MP2} and
fitted using Eq. (\ref{eq:Tsallis}). $(b)$ The ratio {\it
data/fit} for the results presented in $(a)$.} \label{Fig_taua}
\end{figure}
In this variable all data lie on a single curve,
\begin{equation}
\frac{dN}{dp^2_T} = \frac{1}{Q^2_0}F(\tau), \label{eq:univerasal}
\end{equation}
with $F(\tau)$ being some energy independent universal function
(cf. \cite{MP2} for details) and $\lambda$ is a parameter.
Actually, to get good agreement with all available data, $\lambda$
has to depend on $p_T$. The best fit, see Fig. \ref{Fig_taua}, is
obtained with $\lambda = \lambda_{eff}(Q) = 0.13 + 0.1
\left(Q^2/10\right)^{0.35}$, where $Q = 2p_T$ (cf., Eq. (11) of
\cite{MP2}). The variable $p'_T$ used in Fig. \ref{Fig_taua} was
obtained by demanding that $p_T$ at energy $W$ should be connected
with $p'_T$ at energy $W'$ via the following relation (cf.
\cite{MP2} for details),
\begin{equation}
p'_T = p_T\left( \frac{W'}{W}\right)^{\frac{\lambda}{\lambda +
2}}. \label{eq:lambda_prim}
\end{equation}
Fig. \ref{Fig_taua} shows the corresponding results together with
a Tsallis fit performed using Eq. (\ref{eq:Tsallis}) with $C =
14.0$, $q = 1.121$, $T = 0.18$ GeV.

Here we would like to check whether already considered data show a
kind of $q$-scaling as seen from the perspective of Tsallis
statistics (and, if so, in what variable). In other words: is it
possible to find, in the framework of the nonextensive statistics,
a variable (different from the $p_T'$ above) which would scale the
$dN/dp_T$ distributions? And, is the parameter $q=1.121$ from the
Tsallis fit in Fig. \ref{Fig_taua}a already universal, or else can
one also have such scaling behaviour for some other value of the
parameter $q$ using a different scaling variable?

\begin{figure}[h]
\begin{center}
\includegraphics[width=0.75\textwidth]{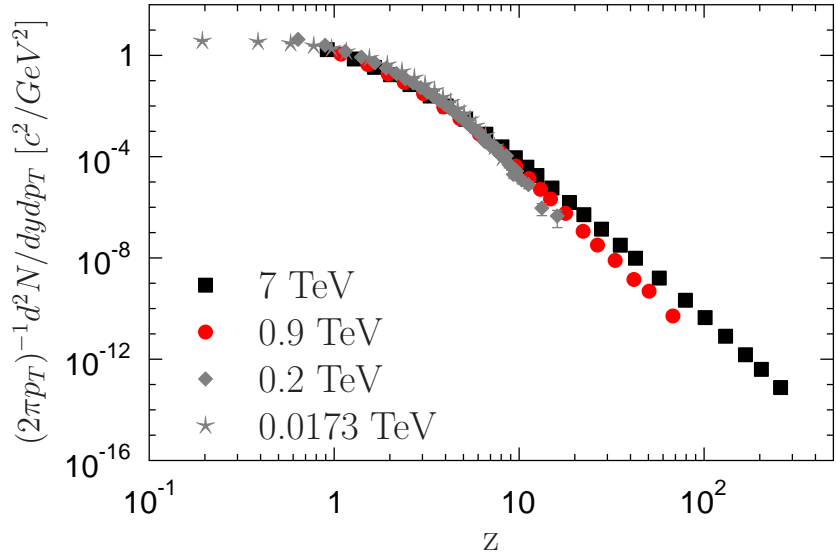}
\caption{(Color online) Data for transverse momentum distributions
for different energies \cite{NA49,UA1,CMS} plotted by using a
scaling variable $z$ as defined in Eq.
(\ref{eq:z}).}\label{Fig_tau}
\end{center}
\end{figure}

\begin{figure}[h]
\includegraphics[width=0.5\textwidth]{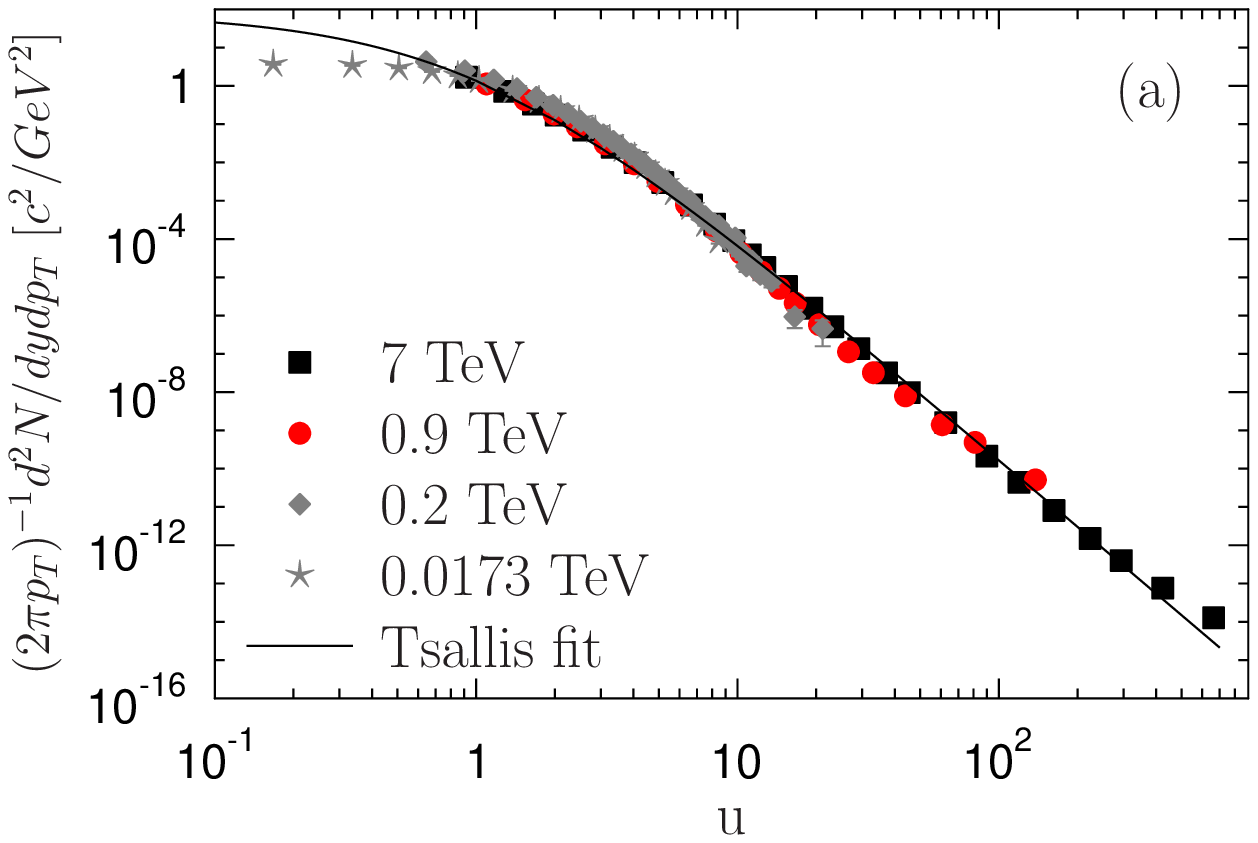}
\includegraphics[width=0.5\textwidth]{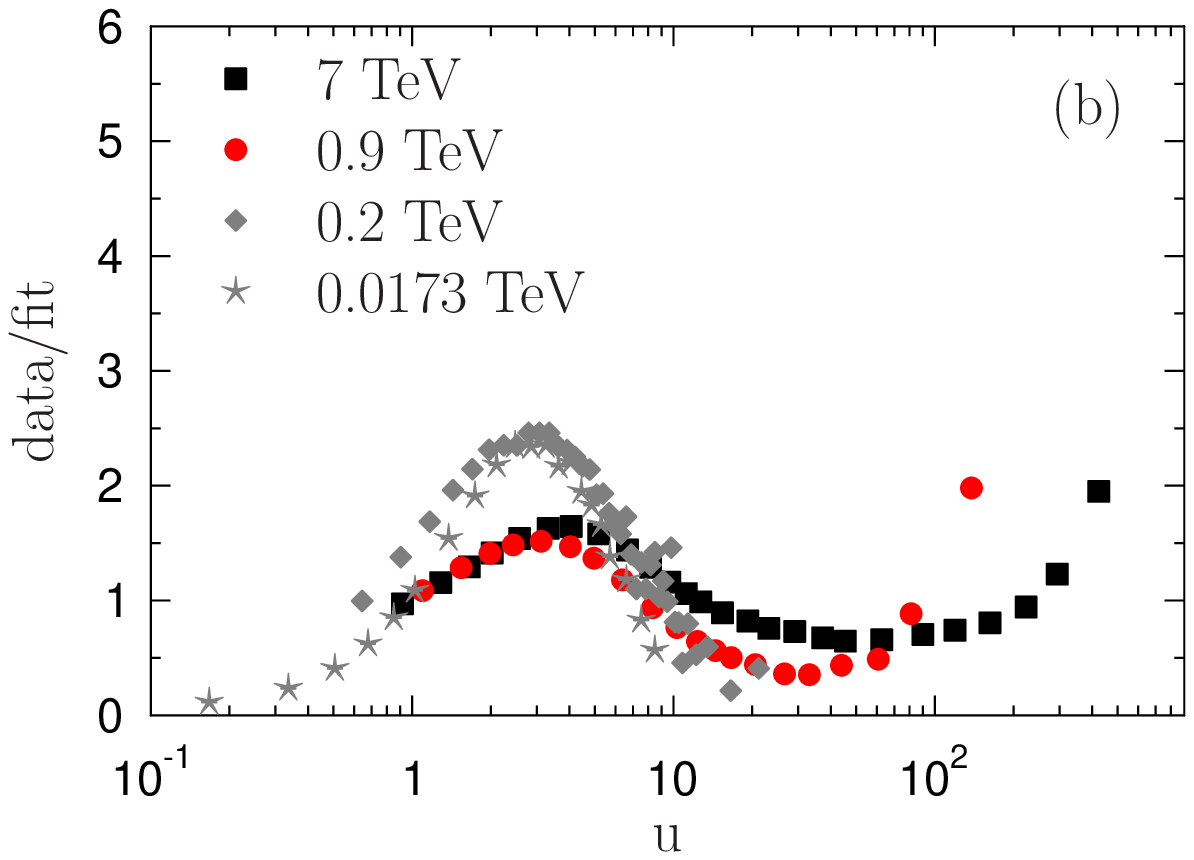}
\caption{(Color online) $(a)$ Data for transverse momentum
distributions for different energies \cite{NA49,UA1,CMS} plotted
by using the scaling variable $u$ defined by Eq. (\ref{eq:u}) and
fitted by Eq. (\ref{eq:Tsallis}) in the variable $u$. $(b)$ The
ratio {\it data/fit} for results presented in $(a)$.}\label{Figu}
\end{figure}

Prompted by KNO scaling as observed in multiplicity distributions
\cite{KNO} we first plot data from Fig. \ref{Fig_exp}a using the
scaled transverse momentum variable,
\begin{equation}
z = \frac{p_T}{\langle p_T\rangle}. \label{eq:z}
\end{equation}
As seen in Fig. \ref{Fig_tau}, already this variable seems to be
nearly satisfactory, except for the largest LHC energies.
Agreement with data can be further improved by using Tsallis
distribution, $h_q(u)$, as given by  Eq. (\ref{eq:Tsallis}), in
which $p_T/T$ is replaced by $u/u_0$, where\footnote{Notice that
$u$ is just a power series in the scaling variable $z$ defined in
Eq. (\ref{eq:z}), $u = \frac{z}{1 - b\cdot z} = \sum_{k =
1}^{\infty} b^{k - 1}z^k = z + bz^2 + b^2 z^3 + \dots,~ $.}
\begin{equation}
u = \frac{p_T}{\langle p_T\rangle - b\cdot p_T} . \label{eq:u}
\end{equation}
Using $\langle p_T(s)\rangle$ taken from an experiment as shown in
Fig. \ref{Fig_ptT}a, and an energy dependent coefficient $b(s) = -
0.0397 + 0.08\left(\sqrt{s}\right)^{-0.075}$, one can reasonably
well fit the data, cf., Fig. \ref{Figu}, with a constant, energy
independent value of $q = 1.172$ (and with $C = 79.4$ and $u_0 =
0.17$). Notice that our result is essentially of the same quality
as that obtained from the geometrical scaling prescription
proposed in \cite{MP2} (cf., Fig. \ref{Fig_taua}).
Both in Fig. \ref{Fig_taua} and in Fig. {\ref{Figu} there are
deviations from Tsallis distributions. Essential here are the
differences between different energies. To show this in both cases
we evaluate the ratios
\begin{equation}
R = \frac{f\left( p_T,\sqrt{s}\right)}{f\left(p_T,\sqrt{s} = 7~
{\rm TeV} \right)}
\end{equation}
of experimental distributions, $f\left( p_T,\sqrt{s}\right) =
\frac{d^3N}{2\pi p_T dp_T dy}$ for different energies $\sqrt{s}$,
which are shown in Fig. \ref{Fig_new6}.
\begin{figure}[h]
\includegraphics[width=0.5\textwidth]{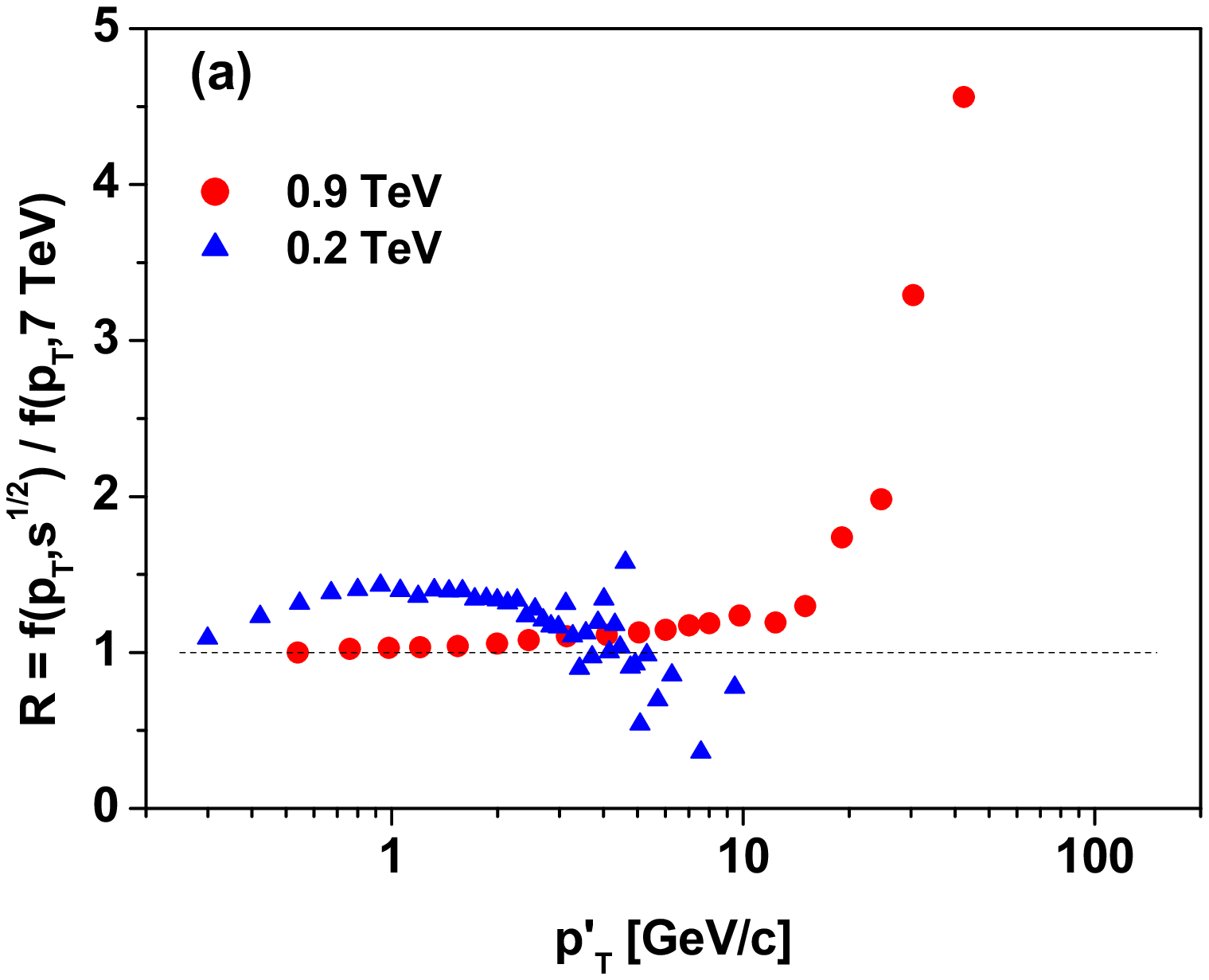}
\includegraphics[width=0.5\textwidth]{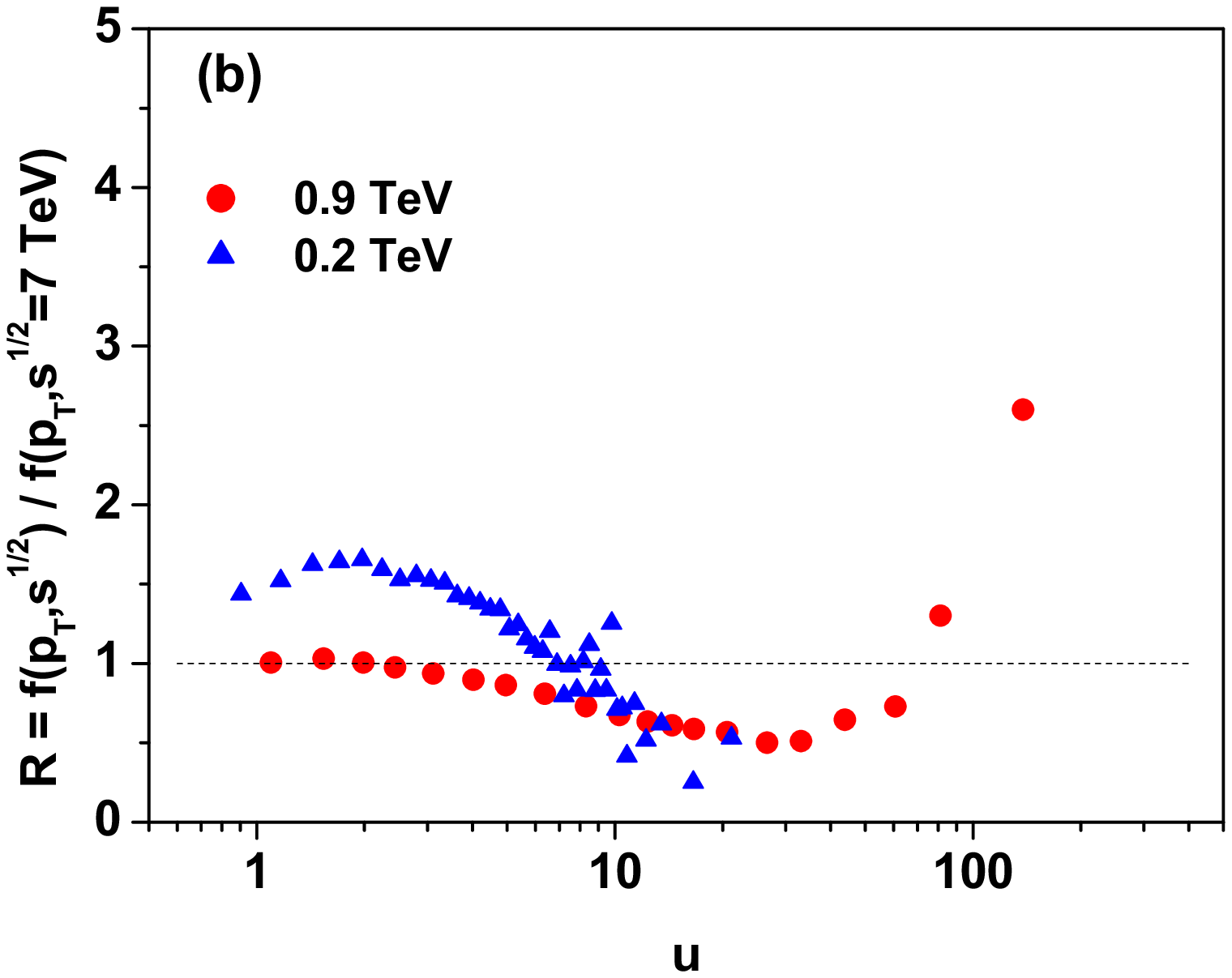}
\caption{(Color online) Ratios of transverse momenta distributions
at energies $\sqrt{s} = 0.9$ and $0.2$ TeV  with reference to the
distribution at $\sqrt{s} = 7$ TeV, expressed in variable $p'_T$
at $(a)$ and variable $u$ at $(b)$.} \label{Fig_new6}
\end{figure}

To justify using the variable $u$ as defined in Eq. (\ref{eq:u}),
note that one can write $u/u_0 = p_T/T_{eff}$ where $T_{eff}$ is
an effective temperature,
\begin{equation}
T_{eff} = T_0 + T_v\left( p_T\right);\quad {\rm with}\quad T_0 =
u_0\cdot \langle p_T\rangle;\quad T_v = - b\cdot u_0\cdot p_T.
\label{eq:Teff}
\end{equation}
This temperature could be related to the possible $p_T$ transfer,
additional to that resulting from a hard collision, perhaps
proceeding by a kind of multiple scattering process, similar, in a
sense, to that proposed on a different occasion in
\cite{multiple}. It is therefore not necessarily connected with
thermodynamics. In fact such $T_{eff}$ also occurs in a
description of the growth of the so called complex free networks
if one associates $p_T$ with the number of links \cite{nets}. When
looking at hadron production from the perspective of stochastic
networks \cite{hlinks} one can argue that the power law seen in
transverse momenta spectra  means that hadronization can be viewed
as a process of formation of some specific network taking place in
the environment of gluons and quark-antiquark pairs formed during
the hadronization process. In this case their actual original
energy-momentum distributions would be of secondary importance in
comparison to the fact that, because of their mutual interactions,
they connect to each other and that this process of connection has
its distinctive dynamical consequences\footnote{The possible line
of reasoning is as follows: Suppose we start with some initial
state consisting of a number $n_0$ of already existing
($q\bar{q}$) pairs (identified with vertices in the network). We
add to them, in each consecutive time step, another vertex (a new
($q\bar{q}$) pair), which can have $k_0$ possible connections
(links in the network language) to the old state. Assume that
quarks are dressed by interaction with surrounding gluons and
therefore "excited" and that each quark interacts with $k$ other
quarks (has $k$ links). Assuming further that the "excitation" of
a quark is proportional to the number of links $k$ (which is
proportional to the number of gluons participating in
"excitation", i.e., existing in the vicinity of a given quark),
the chances to interact with a given quark grow with the number of
links $k$ attached to it. The new links will be preferentially
attached to quarks already having large $k$. This corresponds to
building up a so called preferential network, which evolves due to
the occurrence of new ($q\bar{q}$) pairs from decaying gluons.}.

More formally, following \cite{nets} observe that, whereas
\begin{equation}
\frac{df(x)}{dx} = - \frac{1}{T}f(x)\quad \Longrightarrow\quad
f(x) = \frac{1}{T}\exp\left( -\frac{x}{T}\right), \label{eq:exp}
\end{equation}
the $x$-dependent $T$ in the form,
\begin{equation}
T\rightarrow T(x) = T_0 + (q - 1)x, \label{eq:teff}
\end{equation}
results in a Tsallis distribution:
\begin{equation}
\frac{df(x)}{dx} = - \frac{1}{T_0 + (q - 1)x}f(x)~ \Longrightarrow
~f(x) = \frac{2 - q}{T_0}\left[ 1 - (1 -
q)\frac{x}{T_0}\right]^{\frac{1}{1 - q}}. \label{eq:qexp}
\end{equation}
When fitting data with this distribution, one encounters the
necessity to use in Eq. (\ref{eq:qexp}) the $s$-dependent $q$ and
$T_0$:
\begin{equation}
q = q(s) = q_0 + q'(s)\quad {\rm  and}\quad T_0 = T_0(s) = T'_0 +
T'(s). \label{eq:q0T}
\end{equation}
To compensate for this $s$-dependence, one can modify the
temperature in Eq. (\ref{eq:qexp}), for example by allowing for an
$x$-dependence:
\begin{equation}
T_0 \rightarrow T_{eff}(x) = T'_0 - bx, \label{eq:eff}
\end{equation}
where the parameter $b$ can be $s$-dependent. Returning to Eq.
(\ref{eq:teff}), one now has
\begin{equation}
T(x) = T'_0 + (q - 1)x - bx \label{eq:Tx}
\end{equation}
and that, solving the present form of Eq. (\ref{eq:qexp}), one
finds
\begin{equation}
f(x) = \frac{2 - q + b}{T_0}\left[ 1 - (1 - q +b)\frac{x}{T_0}
\right]^{\frac{1}{1 - q + b}}. \label{eq:qexpprim}
\end{equation}
Identifying now: $xb(s) = T'(s)$ and $b(s) = q'(s)$, one obtains
an energy independent distribution
\begin{equation}
f(x) = \frac{2 - q_0}{T'_0}\left[ 1 - \left (1 - q_0\right)
\frac{x}{T'_0} \right]^{\frac{1}{ 1- q_0}}. \label{eq:sindep}
\end{equation}
In reality the situation is more complicated because here the
$x$-dependence was introduced to $T$ on the level of the
distribution function, not in the differential equation.
Nevertheless, it seems that with such manipulations one can expect
a distribution of the form
\begin{equation}
f(x) \propto \left[ 1 - \left (1 - q_{eff}\right)
\frac{x}{T_{eff}} \right]^{\frac{1}{ 1- q_{eff}}}
\label{eq:sindepp}
\end{equation}
in which the $s$-dependence should be noticeably reduced. This
leads us to the variable $u$ introduced in Eq. (\ref{eq:u}).

\begin{figure}[h]
\includegraphics[width=0.5\textwidth]{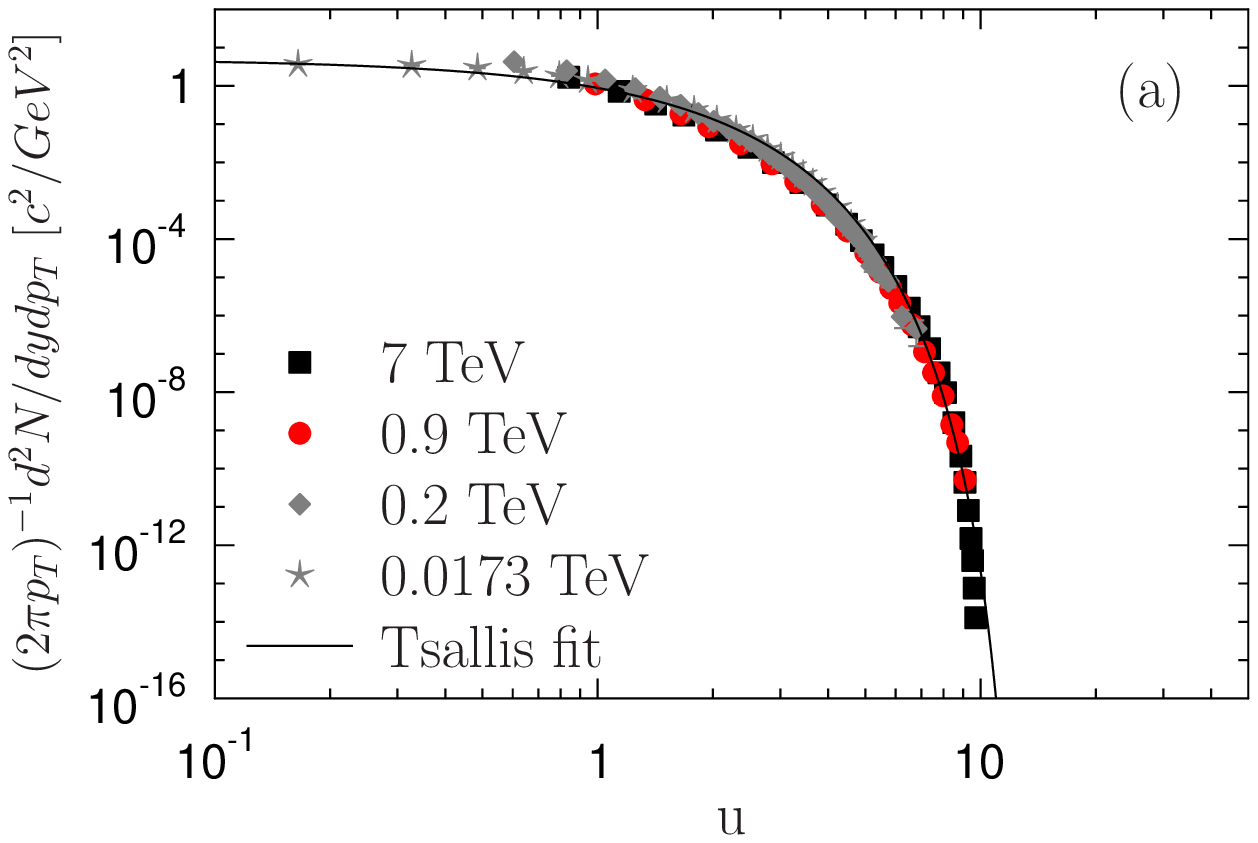}
\includegraphics[width=0.55\textwidth]{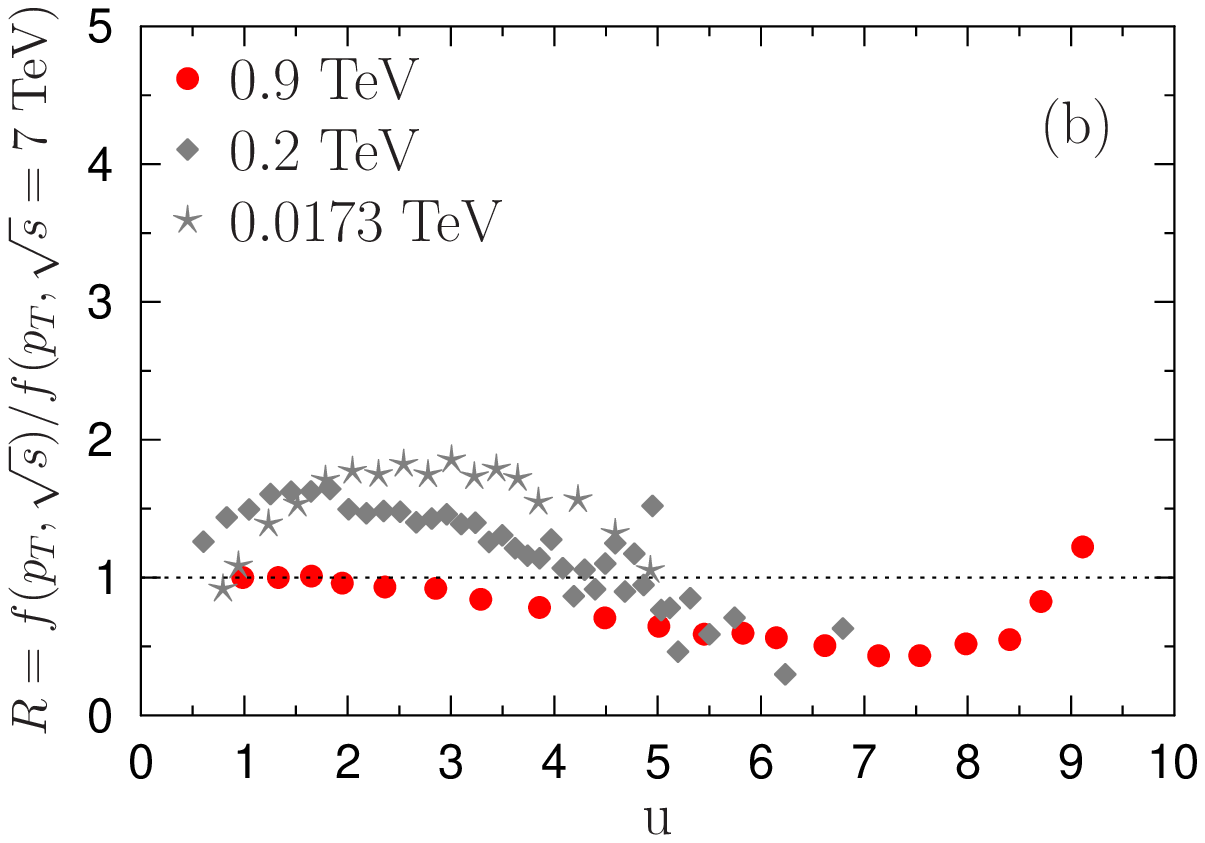}
\caption{(Color online) $(a)$ Data for transverse momentum
distributions for different energies \cite{NA49,UA1,CMS} plotted
by using the scaling variable $u$ defined by Eq. (\ref{eq:u}) with
$b < 0$. $(b)$ Ratios of transverse momenta distributions at
energies $\sqrt{s} = 0.9$, $0.2$ and $0.0173$ TeV  with reference
to the distribution at $\sqrt{s} = 7$ TeV .
 } \label{Fig_new7}
\end{figure}

\begin{figure}[h]
\includegraphics[width=0.5\textwidth]{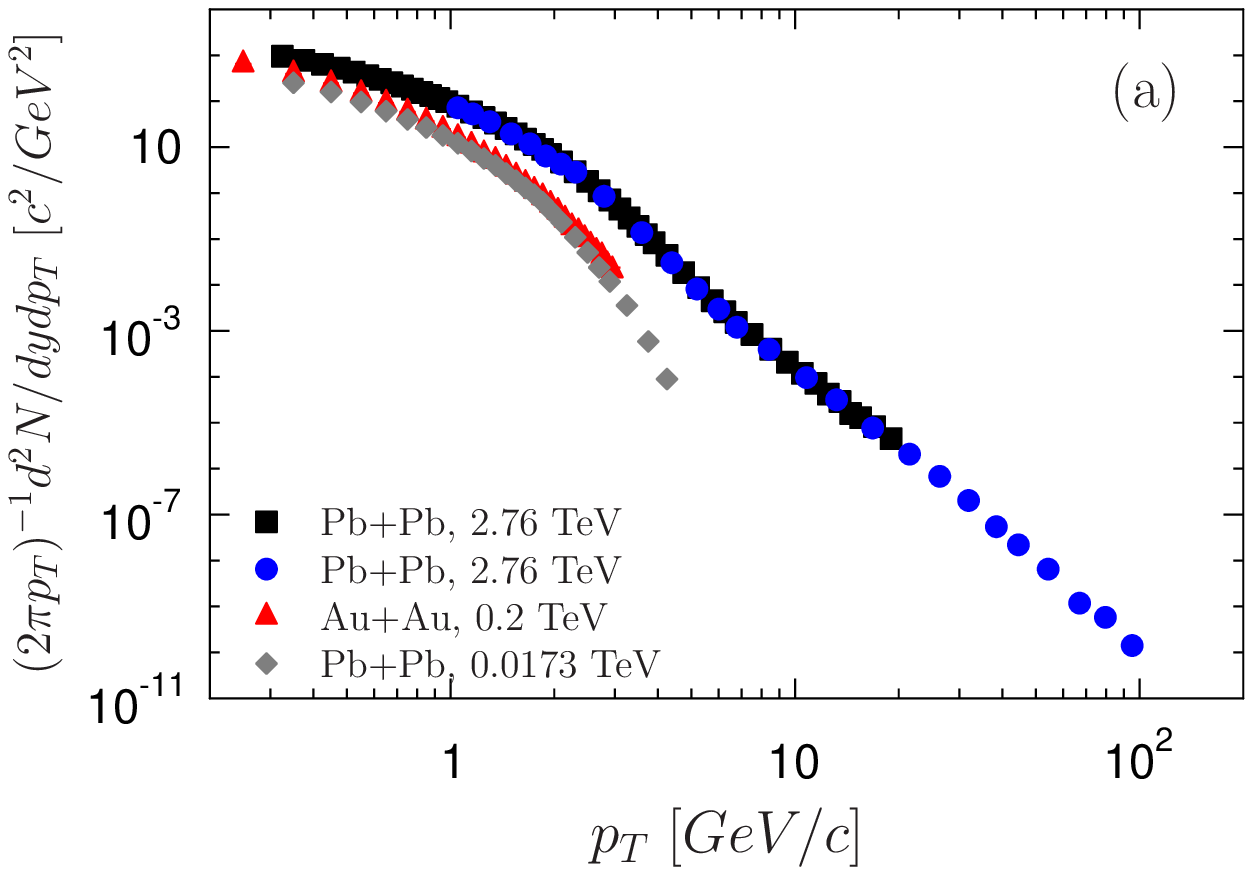}
\includegraphics[width=0.5\textwidth]{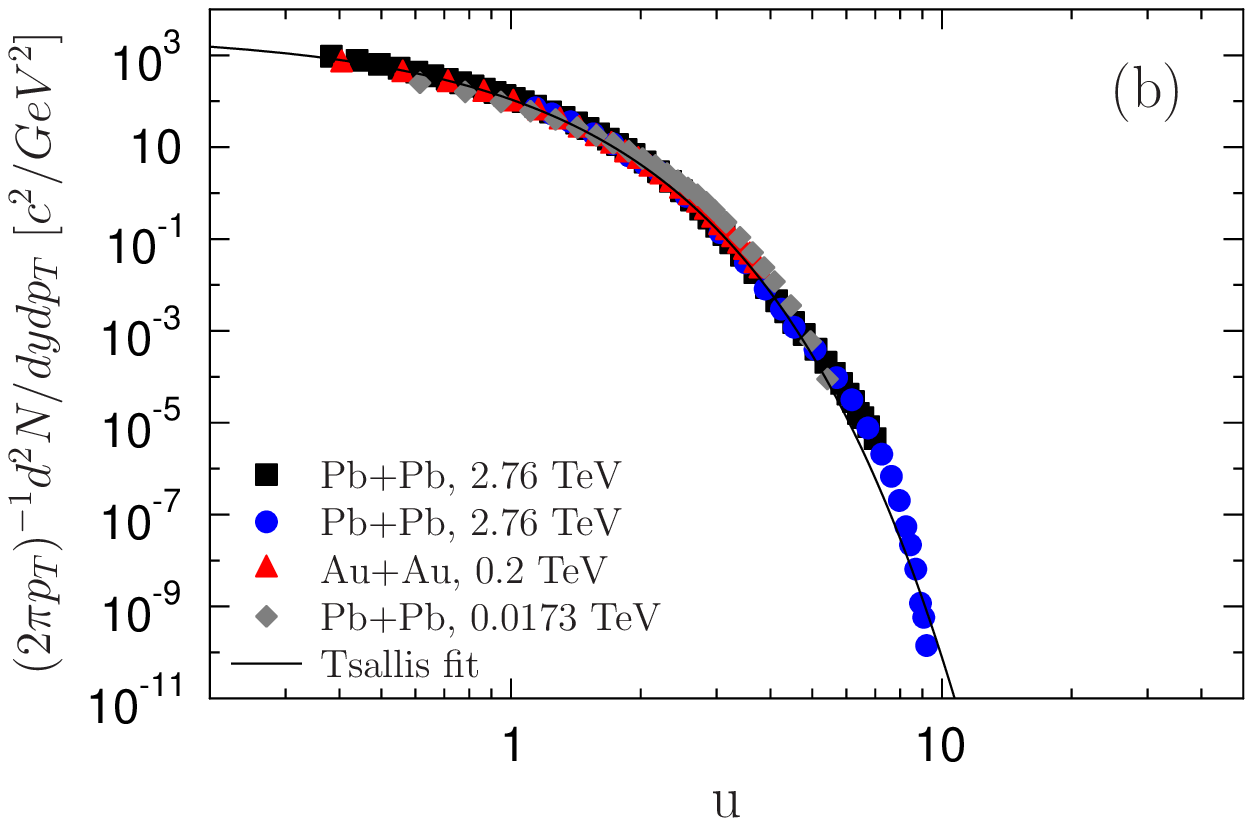}
\caption{(Color online) $(a)$ Data for transverse momentum
distributions for central A+A collisions at different energies.
Data for Pb+Pb at $2.76$ TeV comes from ALICE \cite{ALICE} and CMS
\cite{CMS_AA}, data for Au+Au at $0.2$ TeV come from PHENIX
\cite{PHENIX_AA} and for Pb+Pb at $0.0173$ TeV from NA49
\cite{NA49_AA}. $(b)$ The same plotted by using the scaling
variable $u$ defined by Eq. (\ref{eq:u}) with $b < 0$.
 } \label{Fig_new8}
\end{figure}

Notice that, whereas for the choice of $b(s)$ used above one has
$b > 0$ and $T_{eff}$ was decreasing with $p_T$ (increasing, in
the network approach discussed above, action of the preferential
attachement) one can also choose a parametrization for which $b <
0$ and $T_{eff}$ increases with $p_T$ cancelling effects of
"preferential attachment". For example, for $b(s) = - 0.109 +
0.115 \left(\sqrt{s}\right)^{-0.3}$, one obtains a scaling of
distributions in the variable $u$ with the quasi-exponential form
of the function $h(u)$ (i.e., in a Tsallis distribution with $q$
close to $1$) . The corresponding results are shown in Fig.
\ref{Fig_new7}, together with the corresponding Tsallis fit for $q
= 0.955$ (and $C=5$, $u_0 = 0.6$).

Actually, the same kind of scaling is also possible for $A+A$
collisions for $b(s) = -0.052 -
0.0002\left(\sqrt{s}\right)^{0.7}$. In Fig. \ref{Fig_new8} one can
see distributions of $p_T$ for central collisions ($0-5\%$
centrality) for different data together with the Tsallis fit for
$q=0.9999$ (and $C=2800$, $u_0=0.32$).

To summarize: it is possible to fit all available data on $p_T$
distributions using some universal, energy independent, parameter
$q$. Therefore there is a possibility of $q$-scaling. This can be
done by choosing a variable $u$ defined in Eq. (\ref{eq:u}) in the
distribution $h(u)$ given by Eq. (\ref{eq:Tsallis}). Scaling can
be achieved either by increasing $q$ to $q = 1.172$ (for $b > 0$,
cf., Fig. \ref{Figu} ) or by decreasing it to $q \sim 1$ (for $b <
0$, cf., Fig. \ref{Fig_new7}; in this case Eq. (\ref{eq:Tsallis})
almost coincides with Eq. (\ref{eq:H1}))\footnote{Notice that only
parameter $q$ in Eq. (\ref{eq:Tsallis}), i.e., for variable $p_T$,
has physical sense. In this case it can, for example, be connected
with temperature fluctuations, cf., Eq. (\ref{eq:q}). After
rescaling the variable (i.e., when using $u$ instead of $p_T$) $q$
changes its meaning and can be both greater or smaller than all
the experimentally observed values.}. The observed Tsallis
distributions do not necessarily indicate thermalization of the
system considered (there are numerous examples of non-thermal
sources of Tsallis distributions, cf. \cite{WW2}). The possible
explanation we propose is based on the description of a
hadronization process in analogy with complex free networks
\cite{hlinks}. Alternativelly, one can interpret
Eq.(\ref{eq:qexp}) as a stationary solution of the Fokker-Planck
equation corresponding to a Langevin equation with multiplicative
noise with nonzero mean value $\langle \eta (t) \rangle = 1 - q$
\cite{TSB} (cf. also \cite{WW2}). The possible connection with QCD
based ideas \cite{multiple} is also indicated (but this would
demand special attention, which is outside of the scope of this
paper).

\section*{Acknowledgment}

Acknowledgment: Partial support (GW) of the Ministry of Science
and Higher Education under contract DPN/N97/CERN/2009 is
gratefully acknowledged. We would like to warmly thank Dr Eryk
Infeld for reading this manuscript.

\end{document}